# LATEST DEVELOPMENT OF ELECTROPOLISHING OPTIMIZATION FOR 650 MHz NIOBIUM CAVITY*

V. Chouhan†, D. Bice, D. Burk, S. Chandrasekaran, A. Cravatta, P. Dubiel, G. V. Eremeev, F. Furuta, O. Melnychuk, A. Netepenko, M. K. Ng, J. Ozelis, H. Park, T. Ring, G. Wu, Fermi National Accelerator Laboratory, Batavia, USA
B. Guilfoyle, M. P. Kelly, T. Reid, Argonne National Laboratory, Lemont, USA

## Abstract

Electropolishing (EP) of 1.3 GHz niobium superconducting RF cavities is conducted to achieve a desired smooth and contaminant-free surface that yields good RF performance. Achieving a smooth surface of a large-sized elliptical cavity with the standard EP conditions was found to be challenging. This work aimed to conduct a systematic parametric EP study to understand the effects of various EP parameters on the surface of 650 MHz niobium cavities used in the Proton Improvement Plan-II (PIP-II) linear accelerator. Parameters optimized in this study provided a smooth surface of the cavities. The electropolished cavity showed a significantly higher accelerating gradient meeting baseline requirement and qualified for further surface treatment to improve the cavity quality factor.

## INTRODUCTION

Electropolishing (EP) is a widely utilized technique for treating metal surfaces. It is commonly employed to process niobium made superconducting radiofrequency (SRF) cavities [1–3], which are key components in particle accelerators. These cavities are operated at high frequencies, typically ranging from several hundred megahertz to a few gigahertz, in order to accelerate charged particles. The proton improvement plan-II (PIP-II) linear accelerator (linac), with a target energy of 800 MeV, will incorporate various types of cavities, including low-$\beta$ (0.61) and high-$\beta$ (0.92) 650 MHz elliptical 5-cell cavities. These cavities are known as LB650 and HB650, respectively.

To meet the field gradient specification, the PIP-II elliptical cavities will undergo electropolishing to achieve a smooth interior surface. In a previous study, we emphasized the significance of cathode size by comparing the surface morphology of LB650 cavities. The larger cathode surface area provided required conditions including current plateau for electropolishing, as determined by *I-V* measurements [4]. It was previously established that the standard voltage of 18 V was insufficient for performing electropolishing of the LB650 cavities, particularly with a cathode surface area of approximately 5% of the cavity surface area. Increasing the cathode surface area to approximately 10% significantly reduced the EP voltage. However, it was still higher than the standard 18 V when the cavity temperature was 22 ºC set in bulk EP process [4].

___

* This work was supported by the United States Department of Energy, Offices of High Energy Physics and Basic Energy Sciences under contract No. DE-AC02-07CH11359 with Fermi Research Alliance.
† vchouhan@fnal.gov

In this study, we present a new cathode design specifically designed for electropolishing the HB650 elliptical cavities. The results include cavity's surface and SRF performance, which are also compared to those obtained with LB650 cavities.

## SETUP

### EP setup

The EP process for the 650 MHz cavities was conducted using a horizontal EP tool located at Argonne National Lab. Figure 1 displays a photograph of the EP setup, featuring a horizontally assembled 650 MHz 5-cell cavity.

For the standard EP procedure, a power supply with specifications of 80 V × 500 A was used. Throughout the EP process, the temperature of the cavity was regulated by spraying water onto the exterior surface of the cavity wall. The system utilized in this study was identical to the one used for EP of LB650 cavities. Further information about the system and data logging can be found elsewhere. [4, 5].

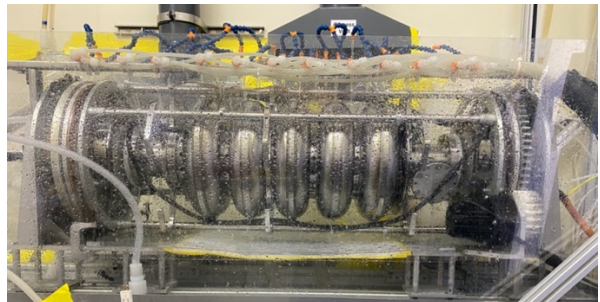

Figure 1: EP tool with a 650 MHz 5-cell cavity.

### Cathode Structure

A novel cathode with an enhanced surface area was developed for EP of HB650 cavities. The surface area of HB650 cavity was 1.951 m$^2$. Fig. 2 illustrates the design of the patented cathode [6]. There was no tape masking applied to the iris and beam tube regions. Instead, a floating mask that was set over the cathode pipe in the iris and beam tube regions was used. The entire cathode surface area was exposed to the acid to maximize the surface area of the cathode. This new design improved the ratio of cathode to anode surface area to approximately 3:10. Additionally, the cathode incorporated a spacious cross-sectional opening, which might facilitate better acid conductance and minimize the buildup of hydrogen gas around the cathode surface.

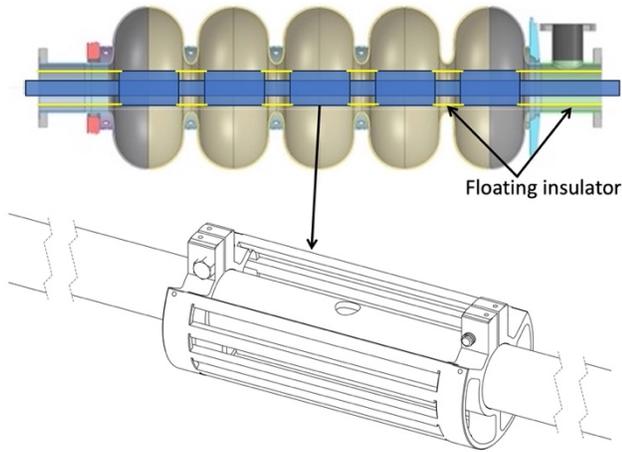

Figure 2: Cavity with a schematic of the cathode (top). Yellow lines show insulating cover over the cathode in iris and beam tube regions. Cathode structure used for EP of the HB650 cavity (bottom) [6].

## RESULTS

### Polarization Curves

The polarization curves were obtained using the novel cathode at various temperatures of the cavity. In this context, the cavity temperature refers to the average surface temperature when the current started plateauing in the corresponding *I-V* curve. Figure 3 illustrates the curves obtained within a temperature range of 10-25 ºC. Figure 3 (b) shows the onset voltage, which represents the voltage at which the current plateau begins, as a function of the cavity temperature. The horizontal error bars represent the standard deviation in the cavity temperature, while the vertical error bar indicates the maximum possible error in estimating the onset voltage. It was observed that the onset voltage increased linearly with the cavity temperature. A similar trend was observed in the case of LB650 cavities using a cathode with a surface area having 10% of the cavity surface area. For comparison, the onset voltage versus cavity temperature curve for LB650 cavities was added in Fig. 3 (b). The novel cathode design resulted in a reduction in the onset voltage by 3 V, providing an advantage over the cathode used in electropolishing LB650 cavities.

### Surface Processing

The surface of an as-received HB650 5-cell cavity (B92F-RI-203) was processed with bulk EP followed by heat treatment at 900 ºC for 3 h and light EP.

EP of the cavity was performed with the novel cathode to evaluate the impact of its high surface area on the cavity surface morphology and its SRF performance. The cavity temperature during bulk and light EP was maintained at 25 ºC and 12 ºC, respectively. The details of the parameters and removal thickness are mentioned in Table 1. Current and temperature profiles during bulk and light EP are shown in Fig. 4. The EP rate was estimated to be ~0.15 and ~0.08 µm/min in bulk and light EP, respectively.

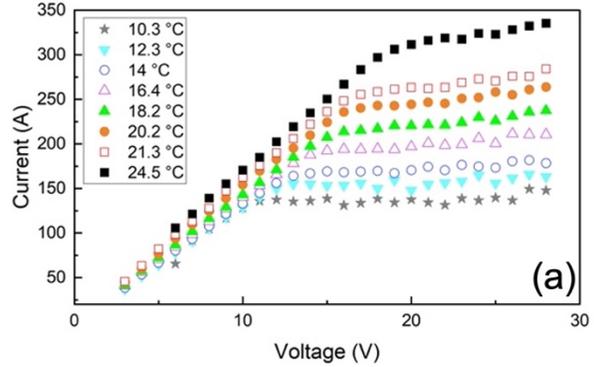

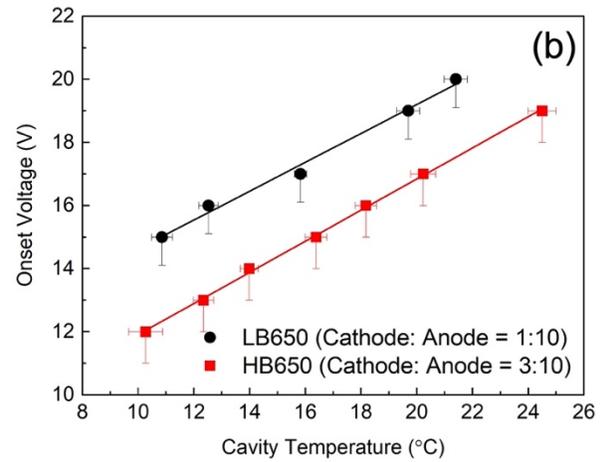

Figure 3: (a) Polarization curves at various cavity temperatures during EP of the HB650 cavity. (b) Onset voltage as a function of cavity temperature for the HB650 and LB650 cavities [3].

The surface after bulk removal was inspected with an optical camera to find the impact of the novel cathode and the applied conditions. An optical image of the equator surface in cell-1 is shown in Fig. 5. The surface was found uniform in all the cells. The surface appeared smooth, and no adverse impact of the large cathode surface area was seen on the surface. The iris and beam tube were also found smooth and pit-free. The surface smoothness was confirmed by making replica of the equator. The replica surface was examined with laser confocal scanning microscope. The surface roughness *Ra* was estimated to be 0.6±0.1 µm.

Table 1: EP parameters applied to HB650 cavity and removal thicknesses in bulk and light EP.

| Parameters | Bulk EP | Light EP |
| --- | --- | --- |
| Voltage (V) | 25 | 22 |
| Acid flow (L/min) | ~8 | ~8 |
| Cavity temperature (ºC) | ~25 | ~12 |
| Cavity rotation (rpm) | ~1 | ~1 |
| Removal thickness (μm) | 160 | 10 |

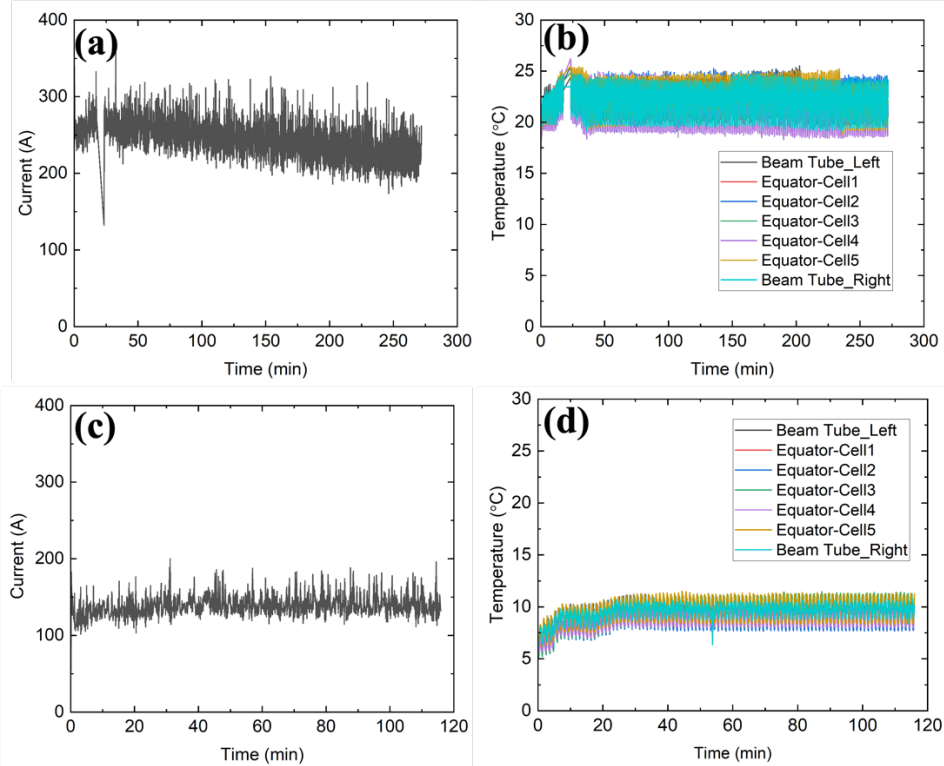

Figure 4: Current and temperature profiles in bulk EP (a and b) and light EP (c and d).

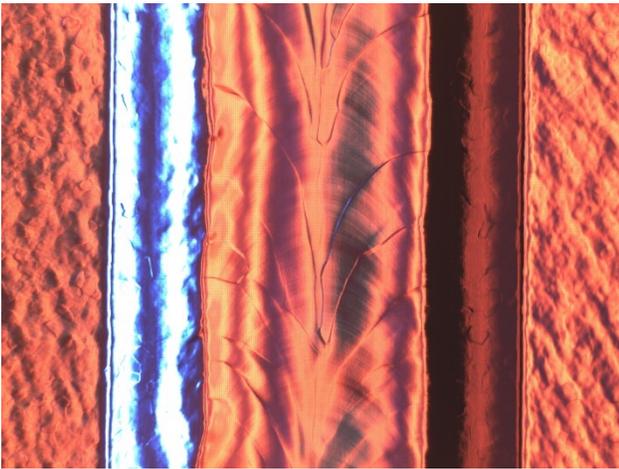

Figure 5: Optical image of the equator surface in the cell-1. The image size is 12 × 9 mm.

### Vertical Test Result

The SRF performance of the cavity was assessed by conducting tests in a vertical cryostat at a temperature of 2 K. The resulting $Q_0$ (quality factor) versus $E_{acc}$ (accelerating gradient) curve is presented in Fig. 6. The cavity was tested up to $E_{acc}$ of 29.6 MV/m at $Q_0$ value of $2.7\times10^{10}$ without experiencing quenching. The performance was limited by field emission. The baseline performance of the cavity met the specification of the cryomodule. In accordance with the surface processing protocol for HB650 cavities in the PIP-II linac, the cavity will undergo a process involving 2/0 N-doping to enhance the $Q_0$ value. Currently, the cavity is awaiting further testing to confirm its performance improvements.

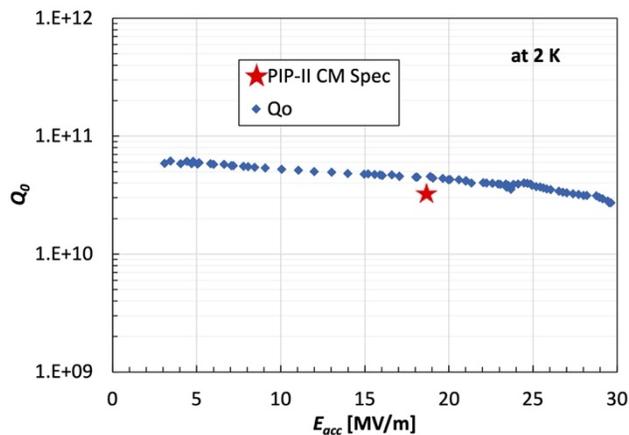

Figure 6: $Q_0$ versus $E_{acc}$ curve showing the cavity's SRF performance. The star-shaped symbol in the plot shows specification of cavity performance in a cryomodule.

## DISCUSSION

The lower onset voltage observed in EP with the novel cathode can be explained by the degree of cathode polarization. The larger surface area reduced the current density and quantity of hydrogen gas bubbles, generated in the chemical reduction reaction, on the cathode surface. Additionally, the cathode design might help in reducing the accumulation of hydrogen gas bubbles, which in turn decreases cathode screening and polarization. The lower cathode current density might result in a lower voltage drop at the cathode-electrolyte interface due to electrical resistance. The reduced cathode screening could also contribute to a lower voltage drop in the novel cathode with a larger surface area. This effect was observed when the cathode surface area was increased from 5% to 10% of the cavity surface area [4]. These results suggested that a large cathode surface is necessary for EP of large-sized cavities such as LB650 and HB650 at a lower voltage.

The reduced voltage requirement for EP can decrease the power load on the cathode and potentially minimize sulfur generation during the EP process. Sulfur typically manifests as particles on the cathode surface. Additionally, the cathode enables EP to be conducted at temperatures above 22 ºC, which was typically used for LB650 cavities, while maintaining a voltage of 25 V to slightly enhance the EP rate in bulk EP. The EP of another HB650 5-cell cavity was successfully demonstrated at 28 ºC and 25 V, resulting in a surface as smooth as the B92F-RI-203. Moreover, the lower onset voltage in EP of nitrogen doped surface reduces the risk of surface pitting, as discussed elsewhere [6]. With the novel cathode, it is possible to perform EP at a low temperature of 12 ºC with an applied voltage of 18 V while maintaining it 5 V higher than the onset voltage. All EP procedures were carried out with the understanding that the applied voltage should be at least 5 V higher than the onset voltage. This criterion was established and enforced due to the likelihood of the beam tube, iris, and the surface near the iris having a lower onset voltage compared to the equator region. These regions might show a greater influence on the $I$-$V$ curve trend than the equator surface. Consequently, the average $I$-$V$ curve in EP might reach a plateau at a lower voltage than that required for the equator surface. An EP study conducted on a 9-cell cavity demonstrated notable differences between the average $I$-$V$ curve and measurements obtained solely from isolated coupons located at the equator, as described in my paper [2].

The smooth surface was achieved as the EP was performed in the current plateau region in the I-V curve. The smooth cavity surface and good performance of the cavity in the vertical test validated that the large surface area of the novel cathode and applied EP parameters were within an optimum range.

## CONCLUSION

A novel cathode, designed with a significantly larger surface area, was developed for the EP process of the HB650 cavity. This innovative cathode design led to a reduction in the onset voltage observed in the $I$-$V$ curves. As a result, EP could be performed at a comparatively lower voltage. The open structure of the cathode also played a role in mitigating the accumulation of hydrogen gas bubbles around the cathode, consequently reducing cathode screening.

To evaluate the effectiveness of the novel cathode, cavity B92F-RI-203 underwent electropolishing using the new cathode and the associated parameters. The electropolishing process yielded a smooth surface, and the cavity exhibited excellent SRF performance. $E_{acc}$ of the cavity reached 29.6 MV/m at a $Q_0 = 2.7 \times 10^{10}$ without experiencing a quench. The surface morphology and SRF performance of the cavity provided validation that both the cathode surface area and the applied parameters fell within an optimal range.

## ACKNOWLEDGEMENTS

This manuscript has been authored by Fermi Research Alliance, LLC under Contract No. DE-AC02-07CH11359 with the U.S. Department of Energy, Office of Science, Office of High Energy Physics.